\RequirePackage{fix-cm}
\documentclass[onecolumn,epjc3]{svjour3}  
\smartqed  
\usepackage{graphicx}
\usepackage{amsmath}
\usepackage{amssymb}
\usepackage{booktabs}
\usepackage{multirow}
\usepackage{color}

\journalname{European Journal of Physics A}

\begin{document}

\title{Measuring two photon exchange in elastic nuclear scattering with $e^+/e^-$ charge asymmetries}

\author{
T.~Kutz\thanksref{mit,gw,e1}
\and
A.~Schmidt\thanksref{gw} 
}

\thankstext{e1}{tkutz@mit.edu}


\institute{
Massachusetts Institute of Technology, Cambridge, MA 02139, USA \label{mit}
\and
George Washington University, Washington, DC 20052, USA \label{gw}
}

\date{Received: date / Accepted: date}

\maketitle

\begin{abstract}
Measurements of nuclear $\beta$-decay are commonly used to extract elements of the quark-mixing Cabibbo–Kobayashi–Maskawa (CKM) matrix.  The precision of these measurements is currently limited by theoretical uncertainties in electroweak radiative corrections (EWRC) arising from so-called box diagrams, which involve the exchange of two gauge bosons.  Two photon exchange (TPE) is the most experimentally accessible of such processes, making it a natural choice for providing constraints on the theoretical frameworks used for calculating EWRC.  The cross section asymmetry between elastic $e^+/e^-$ scattering is directly sensitive to this, as the TPE contribution to the cross section has opposite sign for electrons and positrons.  While charge asymmetry measurements have been performed on proton targets, no such data exists for nuclear targets.  Proposed here is a measurement of the $e^+/e^-$ charge asymmetry on various nuclei relevant to $\beta$-decay measurements used for CKM matrix extraction.  Determining the size of TPE through these processes can provide important constraints on EWRC.           


\end{abstract}

\section{Introduction}

The Cabibbo–Kobayashi–Maskawa (CKM) matrix describes the mixing of quark generations due to the weak interaction.  In the Standard Model, the CKM matrix is a unitary $3\times 3$ matrix.  This condition leads to relations between the matrix elements $V_{ij}$, the mixing strength between quark flavors $i$ and $j$.  For example, first-row unitarity demands that
\begin{equation}
|V_{ud}|^2+|V_{us}|^2+|V_{ub}|^2=1.
\end{equation}
An observed violation of this, or any of the other unitarity relations, would suggest that the $3\times 3$ CKM matrix as we know it is incomplete, and that there may be additional generations of particles to discover. For this reason, precision tests of CKM unitarity can be used to search for and place constraints on new physics. 

The CKM matrix elements can be extracted by normalizing reaction rates of (semi)-leptonic reactions, such as nuclear $\beta$-decay, by the Fermi constant $G_F$ (or equivalently, by the muon lifetime) \cite{MarcianoSirlin2005}.  Super-allowed $\beta$ decays ($0^+ \to 0^+$) are especially useful, as at tree level such reactions include only the weak vector current and are proportional to the weak vector coupling $G_V$.  By conserved vector current (CVC), $G_V$ is not renormalized in the nuclear medium.  However, higher-order electroweak radiative corrections (EWRC) can include the weak axial current, which is subject to hadronic renormalization.  

\begin{figure}[t]
\centering
\includegraphics[width=.5\linewidth]{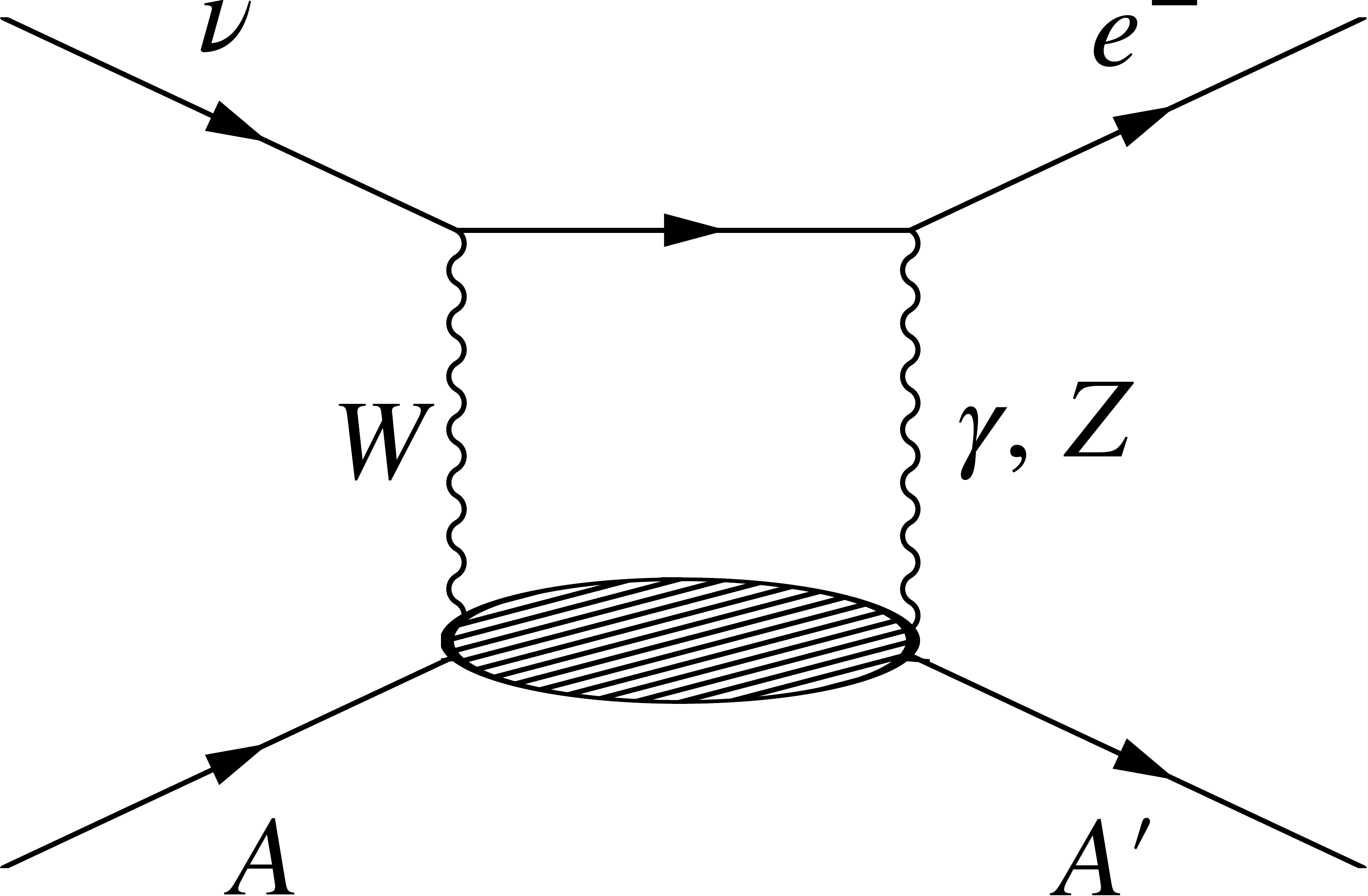}
\caption{The $\gamma W$- and $ZW$-box diagrams contributing to EWRC in nuclear $\beta$-decay.}
\label{fig:feyndiag}
\end{figure}

Of particular interest is the largest first-row element, $V_{ud}$.  The most precise determinations of $V_{ud}$ have come from reaction rates of super-allowed $\beta$-decays:     
\begin{equation}
\left|V_{ud}\right|^2=\frac{G_V^2}{G_F^2} =\frac{2984.432(3)\,\text{s}}{ft(1+\delta'_R)(1+\delta_{NS}-\delta_C)(1+\Delta_R^V)}
\label{eq:vud}
\end{equation}
The $ft$ value is measured experimentally, where $f$ and $t$ are the statistical rate function and partial half-life of the observed decay, respectively.  $\delta_C$ represents an isospin symmetry breaking correction, which will not be discussed here (see for instance \cite{ISB_TH,ISB_Miller}).  $\delta'_R$, $\delta_{NS}$, and $\Delta_R^V$ are terms arising from radiative corrections.  The primary processes to be considered are bremsstrahlung, and the $\gamma W$- and $ZW$-box diagrams depicted in Fig.~\ref{fig:feyndiag}.  

In general, the radiative corrections can be divided into short-distance (high energy) or long-distance (low energy) processes.  The long-distance corrections include bremsstrahlung ($\delta'_R$) and the low-energy contribution to the $\gamma W$-box diagram ($\delta'_{NS}$).  Calculating these corrections requires a model of nuclear structure, and therefore these corrections depend on the specific nucleus involved in the $\beta$-decay process.  The short-distance corrections ($\Delta_R^V$) include the $ZW$-box diagram and low-energy contribution to the $\gamma W$-box diagram.  These calculations are performed using free-quark Lagrangians, and are therefore approximately insensitive to nuclear structure.

The nucleus-dependent corrections can be applied to obtain a ``corrected" value $\mathcal{F}t = ft(1+\delta'_R)(1+\delta_{NS}-\delta_C)$.  Then, Eq.~\ref{eq:vud} becomes
\begin{equation}
\left|V_{ud}\right|^2=\frac{G_V^2}{G_F^2} =\frac{2984.432(3)\,\text{s}}{\mathcal{F}t(1+\Delta_R^V)}
\label{eq:vud2}
\end{equation}
where $\mathcal{F}t$ should be constant regardless of which nucleus is involved in the $\beta$-decay.  The nuclear independence, within uncertainties, of $\mathcal{F}t$ has been observed in nuclei from $A = $ 10 to 74 \cite{HardyTowner2015}.

Currently, precision extractions of $V_{ud}$ are limited by theoretical uncertainties in the calculation of radiative correction \cite{HardyTowner2015}.  In addition to the $\gamma W$- and $ZW$-box diagrams previously discussed, some processes (such as electron scattering) may also have contributions from a $\gamma\gamma$-box diagram, also known as two-photon exchange (TPE).  While TPE is not directly applicable to radiative corrections for $\beta$-decay, any theoretical framework used for calculating EWRC should be able to calculate the TPE diagram.  TPE is easier to access experimentally than interactions involving the $W$ or $Z$ bosons, making it a natural choice for benchmarking theoretical approaches to EWRC.

A number of experimental observables are directly sensitive to the magnitude of TPE.  One such observable is the charge asymmetry, the ratio of the positron to electron elastic scattering cross sections.  As TPE contributes to the elastic scattering of electrons and positrons with opposite sign, the charge asymmetry is sensitive to the size of the effect:
\begin{equation}
R = \frac{\sigma(e^+)}{\sigma(e^-)} \approx \frac{1-2\delta_{2\gamma}}{1+\delta_{even}}
\end{equation}

In this expression, the ratio $R$ has been corrected for the interference term between electron and nuclear bremsstrahlung.  $\delta_{even}$ is the charge-even radiative correction to the ratio.  $\delta_{2\gamma}$ is the correction factor arising from TPE. 

\section{Previous measurements}

Measurements of the ratio of positron-proton and electron-proton elastic cross sections were performed as early as the 1960s \cite{Yount:1962aa,Anderson:1966zzf,Bartel:1967aa,Bouquet:1968aa,Browman:1965zz,Cassiday:1967aa,Mar:1968qd}.  More recent, high-precision measurements have been performed by CLAS \cite{epratio_CLAS}, VEPP-3 \cite{epratio_VEPP}, and OLYMPUS \cite{epratio_DESY}.  The data is highly concentrated in regions of low momentum transfer $Q^2$ and large $\epsilon$ ($>$ 0.7), where TPE is expected to contribute $\lessapprox$ 1\% to the cross section.  A goal of future proton studies will be to extend the kinematics of charge asymmetry measurements into higher $Q^2$ and lower $\epsilon$, where TPE contributions are predicted to reach up to 10\%.  

Similar measurements on nuclear targets do not exist.  Theoretical studies of the TPE contribution to elastic scattering from deuterium \cite{D_kobushkin,D_dong} and helium-3 \cite{He3_blunden,He3_kobushkin} suggest effects of $\mathcal{O}(\lessapprox1\%)$ for $Q^2<1$ GeV$^2$ at large $\epsilon$.     

\section{Proposed measurement}

A positron source that could inject positrons into Jefferson Lab's CEBAF accelerator would make possible, for the first time, measurements of two-photon exchange in elastic nuclear scattering.  We propose a measurement of the elastic $e^+/e^-$ charge asymmetry in various nuclei relevant to constraining EWRC limiting extractions of $V_{ud}$ from super-allowed $\beta$-decays.  Our proposed measurement would require the use of either the high-resolution spectrometers (HRS) in Hall A \cite{HallAnim}, or the high-momentum and super-high-momentum spectrometers (HMS, SHMS) in Hall C, in order to isolate the elastically scattered electrons and positrons.  The specifications of these spectrometers are shown in Table~\ref{tab:spect}.

\begin{table}[t]
\centering
\begin{tabular}{@{}cccc@{}}
\toprule
Spectrometer  & 
\begin{tabular}[c]{@{}c@{}} Resolution \\ ($\delta p/p$)\end{tabular} & 
\begin{tabular}[c]{@{}c@{}}Minimum\\ momentum (GeV)\end{tabular} &
\begin{tabular}[c]{@{}c@{}}Acceptance \\ (msr)\end{tabular} \\ \midrule
HRS (Hall A)  & $2\times10^{-4}$    & 0.8  & 6     \\
HMS (Hall C)  & $8\times10^{-4}$    & 0.5  & 6     \\
SHMS (Hall C) & $1\times10^{-3}$    & 2.0  & 5     \\ \bottomrule
\end{tabular}
\caption{Key parameters of the Jefferson Lab spectrometers that could potentially be used for the measurement.}
\label{tab:spect}
\end{table}

All of these spectrometers have small acceptances with solid angles of a few millisteradians.  As such, choice of spectrometer offers no significant advantage in expected data rates.  The more important considerations are the resolution and minimum momentum of each spectrometer, which will dictate which nuclear targets offer feasible measurements.    

In the following, the maximum beam current for both electrons and positrons is assumed to be 1 $\mu$A.  As this measurement is fully unpolarized, limitations on the maximum polarization of the positron beam or target are not considered. 

\subsection{Nuclear targets}

For the purposes of constraining EWRC to $V_{ud}$, the most useful charge asymmetry measurements are on the daughter nuclei of super-allowed $\beta$-decays used for $V_{ud}$ extraction.  The daughter nucleus for twelve of the best known super-allowed transitions are shown in Table~\ref{tab:nuclei} \cite{hardytowner2005}, along with the natural abundance and first excited state energy of each nucleus.  The latter quantity will determine which nuclei would allow feasible elastic measurements given the resolution of available spectrometers.  

\begin{table}[b]
\centering
\begin{tabular*}{0.8\columnwidth}{@{\extracolsep{\fill}}ccc@{}}
\toprule
Nucleus   & Abundance (\%) & $E^*$ (keV) \\ \midrule
$^{10}$B  & 19.9           & 718.38     \\
$^{14}$N  & 99.6           & 2312.80    \\
$^{22}$Na & trace          & 583.05      \\
$^{26}$Mg & 11.01          & 1808.74     \\
$^{34}$S  & 4.25           & 2127.56    \\
$^{34}$Cl & N/A            & 146.36      \\
$^{38}$Ar & 0.0629         & 2167.47    \\
$^{42}$Ca & 0.647          & 1524.71     \\
$^{46}$Ti & 8.25           & 889.29     \\
$^{50}$Cr & 4.345          & 783.31      \\
$^{54}$Fe & 5.845          & 1408.19     \\
$^{74}$Kr & N/A            & 455.61      \\ \bottomrule
\end{tabular*}
\caption{Daughter nuclei for twelve of the best-known super-allowed $\beta$-decays, with the isotopic abundance and first excited state energy listed for each.}
\label{tab:nuclei}
\end{table}

Of these nuclei, $^{14}$N, $^{26}$Mg, $^{34}$S, $^{38}$Ar, $^{42}$Ca, and $^{54}$Fe are notable for their natural abundance and/or relatively large first excited state energy.  The ability of the spectrometers to resolve the elastic peak is dependent on the initial beam energy.  At 2 GeV, the minimum central momentum allowed by the SHMS, the resolution of the Hall C spectrometers only allow measurements of $^{14}$N, $^{34}$S, and $^{38}$Ar.  At a lower beam energy of 1 GeV, measurements of all 6 nuclei would be possible with both the HRS and HMS.


The most convenient approach would be to design a target ladder capable of holding a variety of gas and solid targets.  For nitrogen and argon, a pressurized 25 cm long aluminum cell with thin entrance and exit windows could be used to contain the target gas.  Similar gas targets have been successfully implemented for electron scattering in the past, notably the argon gas target for experiment E12-14-012 at Jefferson Lab \cite{argon}.  For magnesium, sulfur, calcium, and iron, solid targets with thicknesses of 6 mm or less could be implemented.  Similar solid targets were used for PREX and CREX, which measured elastic scattering from lead and calcium.  As is standard for such target ladders, additional targets for optics and background measurements could be included.  

Based on previous implementations of gas and solid targets at JLab, a nominal target density of 1 g cm$^{-2}$ is assumed.  Combined with the previously mentioned maximum positron beam current of 1 $\mu$A, the achievable luminosities range from 10$^{38}$-10$^{39}$ cm$^{-2}$ s$^{-1}$, depending on nuclear species.

\begin{figure}[b]
\centering
\includegraphics[width=0.8\columnwidth]{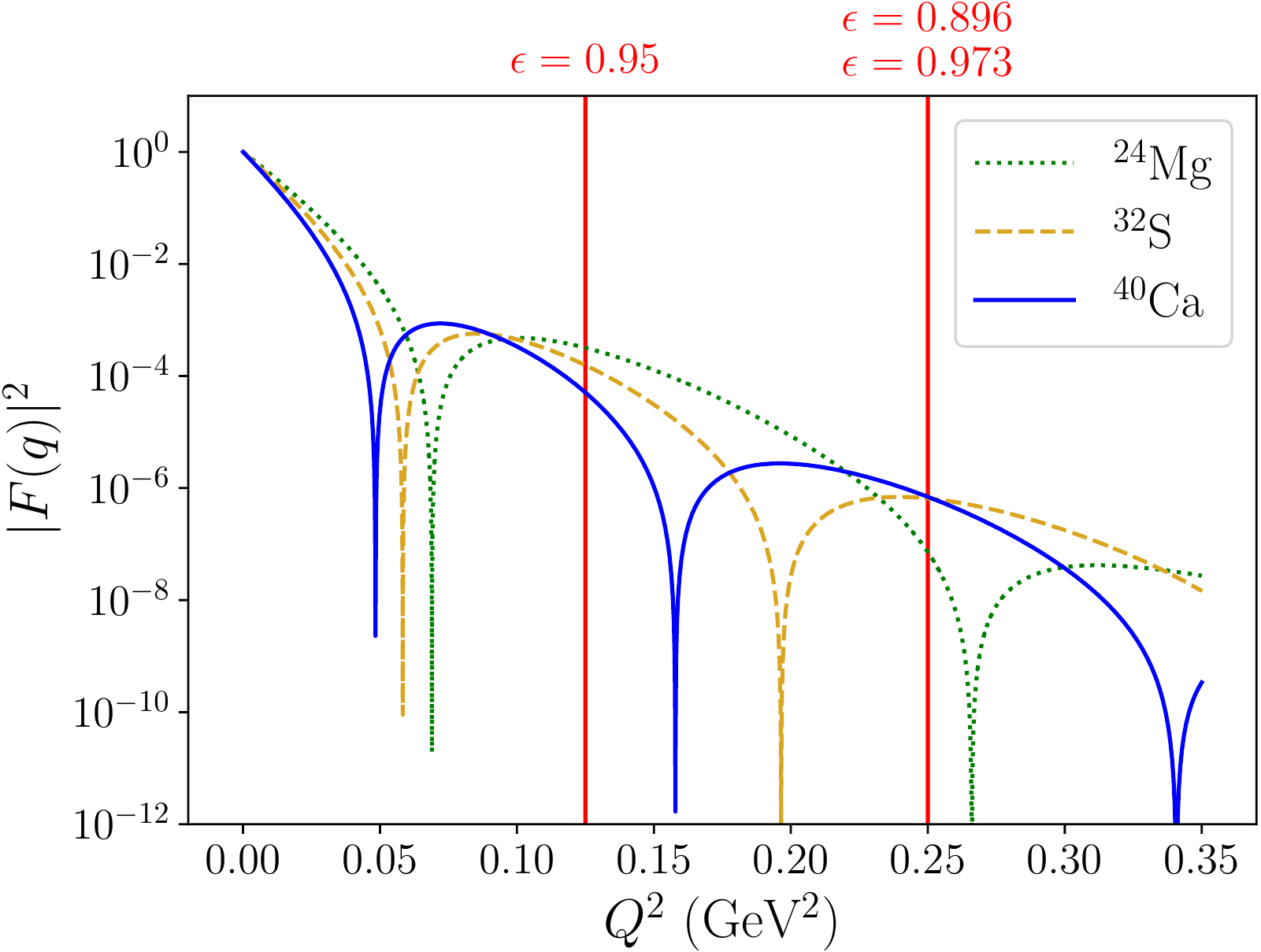}
\caption{Select form factors used for rate estimates.  Red vertical lines indicate the proposed kinematic settings.}
\label{fig:ffs}
\end{figure}

\begin{table}[ht!]
\centering
\begin{tabular}{cccccc}
\hline
$E$ (GeV)            & $\theta_{e^\pm}$ ($^\circ$) & $Q^2$ (GeV$^2$)         & $\epsilon$             & Nucleus        & Days \\ \hline \\[-9pt]
\multirow{6}{*}{1.1} & \multirow{6}{*}{18.12}         & \multirow{6}{*}{0.120} & \multirow{6}{*}{0.950} & $^{14}$N      & 0.5    \\
                     &                             &                         &                        & $^{26}$Mg      & 0.5    \\
                     &                             &                         &                        & $^{34}$S       & 0.5    \\
                     &                             &                         &                        & $^{38}$Ar      & 0.5    \\
                     &                             &                         &                        & $^{42}$Ca      & 0.5    \\
                     &                             &                         &                        & $^{54}$Fe      & 0.5    \\ \hline \\[-9pt]
\multirow{6}{*}{1.1} & \multirow{6}{*}{26.27}      & \multirow{6}{*}{0.250} & \multirow{6}{*}{0.896} & $^{14}$N        & 0.5    \\
                     &                             &                         &                        & $^{26}$Mg      & 7.0    \\
                     &                             &                         &                        & $^{34}$S       & 3.5    \\
                     &                             &                         &                        & $^{38}$Ar      & 2.0    \\
                     &                             &                         &                        & $^{42}$Ca      & 1.5    \\
                     &                             &                         &                        & $^{54}$Fe      & 3.5    \\ \hline \\[-9pt]
\multirow{6}{*}{2.2} & \multirow{6}{*}{13.05}      & \multirow{6}{*}{0.250} & \multirow{6}{*}{0.973} & $^{14}$N        & 0.5    \\
                     &                             &                         &                        & $^{26}$Mg      & 0.5    \\
                     &                             &                         &                        & $^{34}$S       & 1.5    \\
                     &                             &                         &                        & $^{38}$Ar      & 0.5    \\
                     &                             &                         &                        & $^{42}$Ca      & 0.5    \\
                     &                             &                         &                        & $^{54}$Fe      & 0.5    \\ \hline
                     &                             &                         &                        & \textbf{TOTAL} & 25    \\ \hline
\end{tabular}
\caption{Proposed run plan for charge asymmetry measurements from six nuclei at three kinematic settings.}
\label{tab:runplan}
\end{table}

\subsection{Run Plan}

Given the discussion in the previous section, we propose a measurement of the charge asymmetry on six isotopically-pure nuclear targets: $^{14}$N, $^{26}$Mg, $^{34}$S, $^{38}$Ar, $^{42}$Ca, and $^{54}$Fe.  The runplan is shown in Table~\ref{tab:runplan}.  Measurements are proposed at three different kinematic settings, all of which have $Q^2 \leq 0.250$ GeV$^2$ and $\epsilon \geq 0.9$.  This experimental program  prioritizes measurements of multiple nuclear targets over covering large regions of phase space.  While it may be preferable to perform the measurements at kinematics where the effects of TPE are expected to be larger, the rapid decrease of nuclear form factors with $Q^2$ limits the phase space accessible without prohibitively long beam time.    

To perform estimates of the expected experimental rates, a Monte Carlo simulation of a high-resolution, small-acceptance spectrometer was used.  The simulation employed cross sections calculated from I. Sick's parameterization of nuclear form factors \cite{sick}.  As global data fit parameters \cite{devries} were not available for all nuclei, rates were linearly interpolated to intermediate $Z$ values.  The form factors for three of the six proposed nuclei are shown in Fig.~\ref{fig:ffs}.  Also shown are the kinematic settings proposed in Table~\ref{tab:runplan}.

For each nucleus and kinematic setting, the beam time has been estimated to achieve better than 1\% statistical uncertainty on the $e^+/e^-$ ratio.  It is anticipated that in this era of JLab physics, Hall A will only have one operational HRS.  Further, the kinematic settings listed in Table~\ref{tab:runplan} are incompatible with the SHMS in Hall C.  Therefore, the required beamtime has been calculated based on the rate for a single spectrometer.  A factor of 2 has been included in the beamtime to account for approximately 50\% beam efficiency.

\subsection{Systematics}

As the effect of TPE on the charge asymmetry at the proposed kinematics is expected to be $\lessapprox$ 1\%, control of systematics will be critical to these measurements.  It is a significant benefit that the charge asymmetry is a cross section ratio using the same nuclear target and spectrometer, resulting in the cancellation of many systematic effects.  

Two potentially significant sources systematic uncertainty to consider are:
\begin{enumerate}
\item{To first order, target density normalization cancels in cross section ratios using the same target.  However, density fluctuations in gas targets caused by beam-induced heating (so-called ``target boiling") could differ for electrons and positrons.  This introduces a systematic effect that would need to be understood and corrected.}
\item{Ideally, the spectrometer optics for electrons and positrons would be identical up to a sign.  In practice, this is not always achievable.  Small changes in the optics between spectrometer polarities could create difference acceptances for electrons and positrons.  This systematic effect does not cancel in the ratio.}
\end{enumerate}

Lastly, the measurement is completely unpolarized, and will therefore not be sensitive to uncertainties in polarization measurements.  Even including polarization uncertainty, previous JLab experiments have achieved systematic control necessary to measure parity-violating asymmetries on order of hundreds of parts per billion \cite{PREX,Qweak,PREX2}. Measuring a $\lessapprox$ 1\% charge asymmetry should be achievable.

\section{Summary}

Extractions of the CKM matrix element $V_{ud}$ from nuclear $\beta$-decay measurements are currently limited by theoretical uncertainties in calculating EWRC.  This hinders the use of CKM unitarity as a precision test of the Standard Model.  Two-photon exchange is an experimentally accessible process that can provide a critical benchmark for the theoretical calculation of EWRC.  The charge asymmetry $R = \sigma(e^+)/\sigma(e^-)$ is sensitive to the real contribution of TPE to the elastic cross section.  This proposed 30 day program would complete measurements of the charge asymmetry on a variety of nuclei used for $\beta$-decay extractions of $V_{ud}$.  This would provide a constraint on the EWRC to these processes that could improve the precision of CKM unitarity tests.

\bibliographystyle{spphys}       
\bibliography{references}   

\begin{thebibliography}{10}
\providecommand{\url}[1]{{#1}}
\providecommand{\urlprefix}{URL }
\expandafter\ifx\csname urlstyle\endcsname\relax
  \providecommand{\doi}[1]{DOI \discretionary{}{}{}#1}\else
  \providecommand{\doi}{DOI \discretionary{}{}{}\begingroup
  \urlstyle{rm}\Url}\fi

\bibitem{MarcianoSirlin2005}
W.J. Marciano, A.~Sirlin, Phys. Rev. Lett. \textbf{96}, 032002 (2006).
\newblock \doi{https://10.1103/PhysRevLett.96.032002}.
\newblock
  \urlprefix\url{https://link.aps.org/doi/10.1103/PhysRevLett.96.032002}

\bibitem{ISB_TH}
I.S. Towner, J.C. Hardy, Phys. Rev. C \textbf{77}, 025501 (2008).
\newblock \doi{10.1103/PhysRevC.77.025501}.
\newblock \urlprefix\url{https://link.aps.org/doi/10.1103/PhysRevC.77.025501}

\bibitem{ISB_Miller}
G.A. Miller, A.~Schwenk, Phys. Rev. C \textbf{80}, 064319 (2009).
\newblock \doi{10.1103/PhysRevC.80.064319}.
\newblock \urlprefix\url{https://link.aps.org/doi/10.1103/PhysRevC.80.064319}

\bibitem{HardyTowner2015}
J.C. Hardy, I.S. Towner, Phys. Rev. C \textbf{91}, 025501 (2015).
\newblock \doi{https://10.1103/PhysRevC.91.025501}.
\newblock \urlprefix\url{https://link.aps.org/doi/10.1103/PhysRevC.91.025501}

\bibitem{Yount:1962aa}
D.~Yount, J.~Pine, Phys. Rev. \textbf{128}, 1842 (1962).
\newblock \doi{https://10.1103/PhysRev.128.1842}.
\newblock \urlprefix\url{http://link.aps.org/doi/10.1103/PhysRev.128.1842}

\bibitem{Anderson:1966zzf}
R.~Anderson, B.~Borgia, G.~Cassiday, J.~DeWire, A.~Ito, et~al., Phys.Rev.Lett.
  \textbf{17}, 407 (1966).
\newblock \doi{https://10.1103/PhysRevLett.17.407}

\bibitem{Bartel:1967aa}
W.~Bartel, B.~Dudelzak, H.~Krehbiel, J.~McElroy, R.~Morrison, W.~Schmidt,
  V.~Walther, G.~Weber, Physics Letters B \textbf{25}(3), 242  (1967).
\newblock \doi{http://dx.doi.org/10.1016/0370-2693(67)90055-X}.
\newblock
  \urlprefix\url{http://www.sciencedirect.com/science/article/pii/037026936790055X}

\bibitem{Bouquet:1968aa}
B.~Bouquet, D.~Benaksas, B.~GrossetÃªte, B.~Jean-Marie, G.~Parrour, J.~Poux,
  R.~Tchapoutian, Physics Letters B \textbf{26}(3), 178  (1968).
\newblock \doi{http://dx.doi.org/10.1016/0370-2693(68)90520-0}.
\newblock
  \urlprefix\url{http://www.sciencedirect.com/science/article/pii/0370269368905200}

\bibitem{Browman:1965zz}
A.~Browman, F.~Liu, C.~Schaerf, Phys.Rev. \textbf{139}, B1079 (1965).
\newblock \doi{10.1103/PhysRev.139.B1079}

\bibitem{Cassiday:1967aa}
G.~Cassiday, J.~DeWire, H.~Fischer, A.~Ito, E.~Loh, J.~Rutherfoord, Phys. Rev.
  Lett. \textbf{19}, 1191 (1967).
\newblock \doi{https://10.1103/PhysRevLett.19.1191}.
\newblock \urlprefix\url{http://link.aps.org/doi/10.1103/PhysRevLett.19.1191}

\bibitem{Mar:1968qd}
J.~Mar, B.C. Barish, J.~Pine, D.~Coward, H.~DeStaebler, et~al., Phys.Rev.Lett.
  \textbf{21}, 482 (1968).
\newblock \doi{https://10.1103/PhysRevLett.21.482}

\bibitem{epratio_CLAS}
{CLAS collaboration}, D.~Adikaram, et~al., Phys. Rev. Lett. \textbf{114},
  062003 (2015).
\newblock \doi{https://10.1103/PhysRevLett.114.062003}.
\newblock
  \urlprefix\url{https://link.aps.org/doi/10.1103/PhysRevLett.114.062003}

\bibitem{epratio_VEPP}
I.A. Rachek, et~al., Phys. Rev. Lett. \textbf{114}, 062005 (2015).
\newblock \doi{https://10.1103/PhysRevLett.114.062005}.
\newblock
  \urlprefix\url{https://link.aps.org/doi/10.1103/PhysRevLett.114.062005}

\bibitem{epratio_DESY}
{OLYMPUS Collaboration}, B.S. Henderson, L.D. Ice, D.~Khaneft, C.~O'Connor,
  R.~Russell, A.~Schmidt, J.C. Bernauer, M.~Kohl, et~al., Phys. Rev. Lett.
  \textbf{118}, 092501 (2017).
\newblock \doi{https://10.1103/PhysRevLett.118.092501}.
\newblock
  \urlprefix\url{https://link.aps.org/doi/10.1103/PhysRevLett.118.092501}

\bibitem{D_kobushkin}
A.P. Kobushkin, Y.D. Krivenko-Emetov, S.~Dubni\ifmmode~\check{c}\else
  \v{c}\fi{}ka, Phys. Rev. C \textbf{81}, 054001 (2010).
\newblock \doi{https://10.1103/PhysRevC.81.054001}.
\newblock \urlprefix\url{https://link.aps.org/doi/10.1103/PhysRevC.81.054001}

\bibitem{D_dong}
Y.B. Dong, Phys. Rev. C \textbf{82}, 068202 (2010).
\newblock \doi{https://10.1103/PhysRevC.82.068202}.
\newblock \urlprefix\url{https://link.aps.org/doi/10.1103/PhysRevC.82.068202}

\bibitem{He3_blunden}
P.G. Blunden, W.~Melnitchouk, J.A. Tjon, Phys. Rev. C \textbf{72}, 034612
  (2005).
\newblock \doi{https://10.1103/PhysRevC.72.034612}.
\newblock \urlprefix\url{https://link.aps.org/doi/10.1103/PhysRevC.72.034612}

\bibitem{He3_kobushkin}
A.P. Kobushkin, J.V. Timoshenko, Phys. Rev. C \textbf{88}, 044002 (2013).
\newblock \doi{https://10.1103/PhysRevC.88.044002}.
\newblock \urlprefix\url{https://link.aps.org/doi/10.1103/PhysRevC.88.044002}

\bibitem{HallAnim}
J.~Alcorn, et~al., Nucl. Instrum. Meth. A \textbf{522}, 294 (2004).
\newblock \doi{https://10.1016/j.nima.2003.11.415}

\bibitem{hardytowner2005}
J.C. Hardy, I.S. Towner, Phys. Rev. Lett. \textbf{94}, 092502 (2005).
\newblock \doi{https://10.1103/PhysRevLett.94.092502}.
\newblock
  \urlprefix\url{https://link.aps.org/doi/10.1103/PhysRevLett.94.092502}

\bibitem{argon}
H.~Dai, et~al., Phys. Rev. C \textbf{99}, 054608 (2019).
\newblock \doi{https://10.1103/PhysRevC.99.054608}.
\newblock \urlprefix\url{https://link.aps.org/doi/10.1103/PhysRevC.99.054608}

\bibitem{sick}
I.~Sick, Nuclear Physics A \textbf{218}(3), 509  (1974).
\newblock \doi{https://doi.org/10.1016/0375-9474(74)90039-6}.
\newblock
  \urlprefix\url{http://www.sciencedirect.com/science/article/pii/0375947474900396}

\bibitem{devries}
H.D. Vries, C.D. Jager, C.D. Vries, Atomic Data and Nuclear Data Tables
  \textbf{36}(3), 495  (1987).
\newblock \doi{https://doi.org/10.1016/0092-640X(87)90013-1}.
\newblock
  \urlprefix\url{http://www.sciencedirect.com/science/article/pii/0092640X87900131}

\bibitem{PREX}
{PREX Collaboration}, S.~Abrahamyan, et~al., Phys. Rev. Lett. \textbf{108},
  112502 (2012).
\newblock \doi{https://10.1103/PhysRevLett.108.112502}.
\newblock
  \urlprefix\url{https://link.aps.org/doi/10.1103/PhysRevLett.108.112502}

\bibitem{Qweak}
Q.~Collaboration, D.~Androi{\'{c}}, et~al., Nature \textbf{557}(7704), 207
  (2018).
\newblock \doi{https://10.1038/s41586-018-0096-0}.
\newblock \urlprefix\url{https://doi.org/10.1038/s41586-018-0096-0}

\bibitem{PREX2}
{PREX Collaboration}, D.~Adhikari, et~al.
\newblock An accurate determination of the neutron skin thickness of $^{208}$pb
  through parity-violation in electron scattering (2021)

\end{thebibliography}

%
%

\end{document}